\documentclass[runningheads,a4paper]{llncs}

\usepackage{amssymb}
\usepackage{graphicx}
\usepackage{verbatim}
\usepackage{soul}
\usepackage{hyperref}
\usepackage{cleveref}
\usepackage[caption=false]{subfig}

\usepackage{url}
\urldef{\mailsa}\path|marco.avvenuti@unipi.it|
\urldef{\mailsb}\path|{f.delvigna,c.bacciu,m.petrocchi,a.marchetti,m.tesconi}@iit.cnr.it|
\urldef{\mailsd}\path|paolo.deluca@kcl.ac.uk|
\newcommand{\keywords}[1]{\par\addvspace\baselineskip
\noindent\keywordname\enspace\ignorespaces#1}

\begin{document}

\mainmatter  

\title{Spotting the diffusion of New Psychoactive Substances over the Internet}

\titlerunning{Spotting the diffusion of New Psychoactive Substances over the Internet}

%
%
\author{Fabio~Del~Vigna\inst{1,2}\and Marco~Avvenuti\inst{1}\and Clara~Bacciu\inst{2}\and Paolo~Deluca\inst{3}\and \\Marinella~Petrocchi\inst{2}\and Andrea~Marchetti\inst{2}\and Maurizio~Tesconi\inst{2}}
\authorrunning{F.~Del~Vigna {\it et al.}}

\institute{Dept. of Information Engineering, University of Pisa, Italy\\
\and
Institute of Informatics and Telematics (IIT-CNR), Pisa, Italy
\and
Institute of Psychiatry, Psychology \& Neuroscience, King's College London, UK\\
\mailsa\\
\mailsb\\
\mailsd\\
}

%
%

\toctitle{Lecture Notes in Computer Science}
\tocauthor{Authors' Instructions}
\maketitle

\begin{abstract}
Online availability and diffusion of New Psychoactive Substances (NPS) represents an emerging threat to healthcare systems. In this work, we analyse drugs forums, online shops, and Twitter. By mining the data from these sources, it is possible to understand the dynamics of drug diffusion and its endorsement, as well as timely detect new substances. We propose a set of visual analytics tools to support analysts in tackling NPS spreading and provide a better insight about drugs market and analysis. 
\end{abstract}

\section{Introduction}
\label{sec:intro}

%

Noticeably, health departments of European countries are facing a raising issue: the online trade of substances that lay in a grey area of legislation, known as New Psychoactive Substances (NPS). European Union (EU) continuously monitors the market to tackle NPS diffusion, forbid NPS trade and sensitise people to the harmful effects of these drugs\footnote{\url{http://www.emcdda.europa.eu/start/2016/drug-markets\#pane2/4}; All URLs in the paper have been accessed on July 10, 2016.}. Unfortunately, legislation is typically some steps back and newer NPS quickly replace old generation of substances.

Online shops and marketplaces convey NPS through the Internet~\cite{Schmidt2011}, without any (or with very few) legal consequences. Quite obviously, this attracts drug consumers, which can legally  buy these drugs without risk of prosecution. The risks connected to this phenomenon are high: every year, hundreds of consumers get overdoses of these chemical substances and hospitals have difficulties to provide effective countermeasures, given the unknown nature of NPS. Furthermore, products sold over the Internet with the same name may contain different substances, as well as possible changes in drug composition over time~\cite{Davies2010}.

Social media and specialised forums offer a fertile stage for questionable organisations to promote NPS as a replacement of well known drugs, whose effects have been known for years and whose trading is strictly forbidden. 
Furthermore, forums are contact points for people willing to experiment with new substances or looking for alternatives to some chemicals, but also a discussion arena for those at the first experiences with drugs, as well as trying to stop with substance misuse or looking for advice regarding doses, assumption and preparation.
 
The EU-funded project Cassandra\footnote{\url{http://www.projectcassandra.eu}} investigates the NPS supply chain, lifecycle, and endorsement, through the analysis of popular social media, drug forums, and online shops. Such analysis is vital to timely detect NPS diffusion: this will support governments and health agencies in confining the progress of substance abuse, prohibiting NPS sales and  improving the awareness of citizens towards unhealthy and harmful  behaviours.
%


In this paper, we shed light on the structure and activity of NPS forums and online shops.  
%
The main contributions are as follows: i) we give an insight into two popular forums, Bluelight\footnote{\url{http://www.bluelight.org}}  and Drugsforum\footnote{\url{https://drugs-forum.com}}, hosting drugs discussions since more than one decade; ii) we map NPS sales (as monitored on online shops) and NPS diffusion and distribution (as monitored on discussion forums); and iii) we provide automatic support to timely NPS detection. 

Overall, we show a successful application of Intelligent Data Analysis techniques to complex systems, such as social networks and hierarchical ones. This eases the human exploration and interpretation of the online universe of drugs, with a support for the interactive visualisation of the data analysis results. 

The paper is structured as follows. Next section gives related work in the area. Section~\ref{sec:datasources} gives a panoramic view on our data sources. In Section~\ref{sec:analysis}, we focus on forums and shops structure and activities, by analysing their data and offering a visualisation of the analysis results. Finally, Section~\ref{sec:conc} concludes the paper and gives directions for future work.

\section{Related Work}
\label{sec:relwork}
Recently, academia has started investigating the massive use of social media and online forums to advertise and discuss about psychedelic substances and drugs, and how the preferences of online communities can affect those of consumers. 
Large forums drew attention, being a primary source of information about NPS and a good sample of consumer tastes~\cite{Davey2012}.
Work in~\cite{Ledberg2015} considers the Flashback forum and traces the trend of the discussions, especially in relation with the scheduling of a substance ban. The paper puts in evidence how volumes of discussions drop when a ban is scheduled.
In~\cite{Zawilska2015}, the authors focus on new drugs detection and categorisation by scanning online shops and the dark net. A complete list of the known effects of new drugs, to the publication date, is given in~\cite{Schmidt2011,Hillebrand2010}. 



Small subsets of the contents of the Drugsforum and Bluelight forums, which we deeply analyse in the present paper, have been already considered in~\cite{Soussan2014}, highlighting how large forums embody a cumulative community knowledge, i.e., a stratified knowledge built over years of forum activities, and showing that drugs effects and dosage are among the most discussed topics. 

Other studies explored the abuse of medicines and how these are advertised, e.g., on Twitter, and sold by online pharmacies, with no authorisation~\cite{Katsuki2015,Freifeld2014}. Twitter features a rapid spread of contents, especially through small communities of users, which share common interests and tastes. This is the main reason why it has been investigated to mine patterns of drug abuse, also for non-medical purposes, e.g., to improve students performances in study~\cite{Hanson2013,Hanson2013b}. Furthermore, Twitter allows analysts to comprehend rapid disease diffusion and health issues~\cite{Paul2011}, as well as prices and effects of new drugs~\cite{Oconnor2014}. Nevertheless, social media play an important role also for contrasting the drugs diffusion~\cite{Scott2015} and for preventing end users from further consumption~\cite{Inciardi2010}. Twitter was also extensively mined to detect geographical diffusion of drug consumers over time~\cite{Buntain2015}.


The Web is not the only marketplace where NPS are advertised and sold. Indeed, the TOR network\footnote{\url{https://www.torproject.org}} has drawn much attention from drug consumers and resellers, who search for a channel  to buy and sell drugs that guarantees their anonymity. 
This aspect affects trustworthiness of peers, especially when it is not possible to assess users reputation at all. In~\cite{Hardy2015}, the authors investigated the impact of reputation in Silk Road, one of the most popular marketplaces for drugs in the dark net.
Data analysis often deals with the quality of the results obtained when searching the web. The work in~\cite{Pimpalkhute2014} describes the possibility to improve the recall of queries issued to search engines by exploiting all variants and misspelled words.


With respect to related work, this paper addresses a finer-grained, more detailed picture of NPS data sources and NPS data available on the Internet. As an example, the analysis of forums carried on in~\cite{Soussan2014} was limited in time and quantity. In our work, we overcome this limitation, by analysing more than one decade of data, posted by users all over the world.  Overall, we dealt with more than 4 million and a half posts and more than 500,000 users. Furthermore, we integrated more than one source, by monitoring two forums, Twitter, and a number of online shops. The results of our analysis are conveniently conveyed to the reader via a set of interactive visual web interfaces, which are being integrated into a dashboard that will help researchers mine the wealth of gathered data. Ultimately, we are aligned with recent advances in data analysis leading to applications in pattern mining of, e.g., medical records and human anatomies~\cite{deBono2014,Hielscher2014}.


\section{Data sources}\label{sec:datasources}

This section presents the data sources for our analysis. We collected the data  by developing ad-hoc software, which scrapes websites and uses APIs to crawl social media. 


\subsection{Forums} \label{sec:forums}

Bluelight and Drugsforum are two large forums, which host more than a decade of discussion about drugs and addiction.
Being particularly rich of information, the two forums provide a historical, worldwide background of drug consumption, comprising that related to NPS. Similar to Google Flu Trends\footnote{\url{https://www.google.org/flutrends/about/}} efforts to detect spreading of diseases, the analysis of the forums' content and structure is  significant to understand how psychoactive substances have spread out and to study new infoveillance
strategies, to timely detect drugs abuse.  



The two forums have a hierarchical structure, which enables proper content categorisation. The root of both forums organises content into sub-forums, which can be nested up to several levels of depth. The forums' structures were subject to different content re-organisations over time. 
%

We carried out a Web scraping activity to create a dump of the entire database of discussions from the two forums, following the links between the forums' sections. During the storage phase, we kept track of the forums' hierarchy and structure, maintaining all the tags and metadata associated to each post and thread. Table~\ref{tab:forums} summarises the amount of data available from the two forums.



\subsection{Online shops}
The forums introduced in Section~\ref{sec:forums}  are a primary source of information about drugs reviews, feelings, effects and preparation, but little information is available about the drugs market, such as prices and bulk quantities.
Thus, we focused our attention also on other data sources, dealing with drugs trading. 

Online shops sell both legal and illegal substances. Among others, those that sell NPS have grown in popularity, given the relatively low risks in trading such substances.
Many online shops accept payments in pounds, euros and dollars. Also, bitcoins are often accepted. This opens up the possibility to track price trends and, indirectly, to estimate the popularity and quality (or purity) of drugs. Furthermore, many of the marketplaces are advertised and mentioned on forums and social media. 

We have started an intense scraping activity on a set of online shops to monitor the market availability of different substances.
Online shops can be quite easily found through simple queries to search engines (e.g., ``legal highs" and ``smart drugs"). 
We set up a battery of scrapers
that collect the information that are present on the shops showcases. 
Data is collected on a weekly basis, 
and stored in a relational database, to be easy queryable.
Table~\ref{tab:shops} shows the monitored shops. 

\begin{table}[ht]
    \centering
    \begin{tabular}{|c|c|c|c|c|}
        \hline
        \bf{Forum} & \bf{First post} & \bf{Last post} & \bf{Tot posts} & \bf{Users}\\ 
        \hline
        Bluelight & 22-10-1999 & 09-02-2016 & 3,535,378 & 347,457\\
        Drugsforum & 14-01-2003 & 26-12-2015 & 1,174,759 & 220,071\\
        \hline
    \end{tabular}
    \caption{Drug forums: Posts and Users}
    \label{tab:forums}
    
    \begin{tabular}{|c|c|c|}
        \hline
        \bf{ID} & \bf{Website} & \bf{Substances found} \\
        \hline
        \hline
        1 & \url{http://chem-shop.co.uk} & 7\\
        \hline
        2 & \url{http://researchchemist.co.uk} & 45\\
        \hline
        3 & \url{http://researchchemistry.co.uk} & 56\\        
        \hline
        4 & \url{http://sciencesuppliesdirect.com} & 43\\
        \hline
        5 & \url{http://www.bitcoinhighs.co.uk} & 4\\
        \hline
        6 & \url{http://www.buylegalrc.eu} & 17\\
        \hline
        7 & \url{http://www.legalhighlabs.com} & 33\\
        \hline
        8 & \url{http://www.ukhighs.com} & 51\\
        \hline
        9 & \url{https://www.buyanychem.eu} & 78\\
        \hline
        10 & \url{https://www.iceheadshop.co.uk} & 68\\
        \hline
    \end{tabular}
    \caption{Monitored online shops and number of substances they sell}
    \label{tab:shops}
\end{table}


\subsection{Twitter}
Twitter is extensively used by resellers and ``pharmacies" to advertise psychoactive substances, and by consumers to discuss their effects and share feelings with others~\cite{Freifeld2014,Katsuki2015}.
%
We have collected about 14 million tweets, over the period March 16, 2015 -  February 2, 2016, using the Streaming API\footnote{\url{https://dev.twitter.com/streaming/overview}}, which allows applications to gather tweets in real time fashion. 
We have used a crawler that fetches data relying on a set of ad-hoc keywords. We have also followed a series of Twitter accounts associated to online shops.
In the next section, we will detail the monitored keywords, which we chose among known emerging substances. 

\section{Data analysis and Visualization}
\label{sec:analysis}
This section shows the analysis we have carried out over the data sources described in Section~\ref{sec:datasources}. First, we report on a series of analyses over the two drug forums, with the purpose of figuring out their structural features, how their content is organised, and the geographical distribution of their users. Secondly, we mine the forums textual contents, aiming at looking for new substances mentioned in recent discussions. Finally, we provide a picture on the NPS substances sold on online shops, correlating them with mentions on Twitter and the forums. 





\subsection{Forums: Structural and geographical features}
\begin{figure}[!ht]
  \centering
  \includegraphics[width=0.8\textwidth]{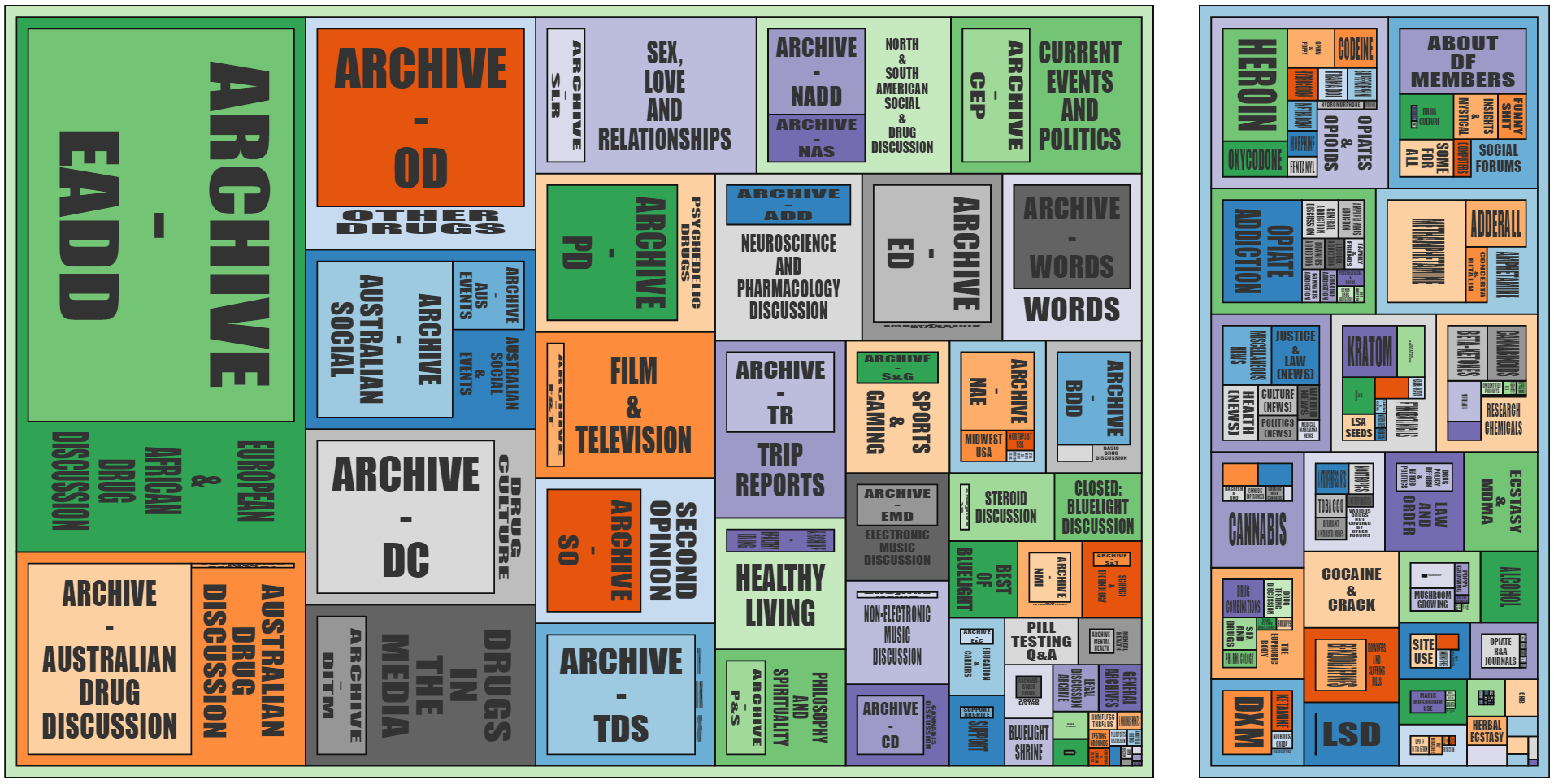}
  \caption{The structure of Bluelight (left) and Drugsforum (right). Bluelight is about three times bigger.}
  \label{fig:treemap}
\end{figure}

To facilitate the investigation of the forums structural features, we have developed a set of visual interfaces. 
Figure~\ref{fig:treemap} depicts the screenshot of a zoomable treemap of the two forums. Nested subsections are represented as nested rectangles, the area of which are proportional to the number of posts a subsection contains. Quick visual comparisons of the forums' size and structure may gather meaningful information. For example, compared to Drugsforum, whose structure is quite complex, Bluelight has a shallow organisation. Also, the names of the subsections suggest that the discussion on Drugsforum is mainly focused on drugs and it follows a rigid categorisation, based on the kind of the substance, while the topics on Bluelight are broader and less related specifically to drugs.

\begin{figure}[!ht]
  \centering
  \includegraphics[width=0.8\textwidth]{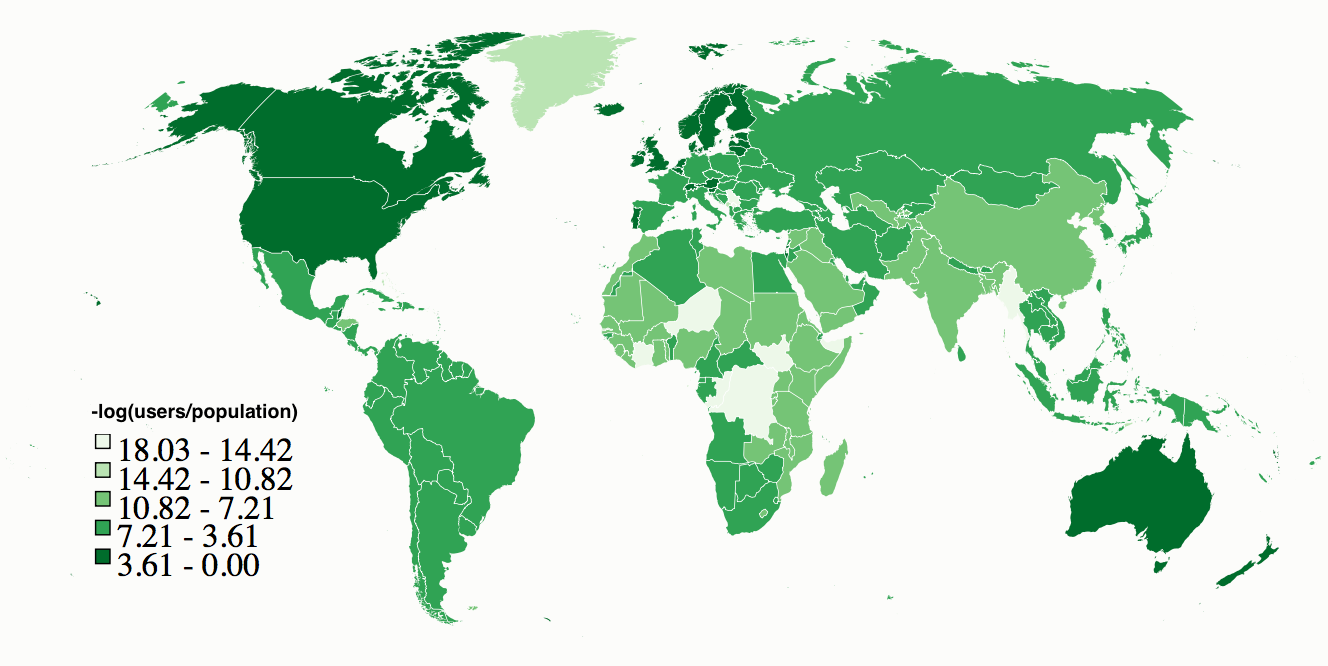}
  \caption{Geographical distribution of Drugsforum users.}
  \label{fig:dfmap}
\end{figure}

Figure~\ref{fig:dfmap} shows the worldwide distribution of the Drugsforum users. The information has been extracted from the users' profiles (when available). Looking at the figure, we understand that drugs discussions on forums is a wide phenomenon, quite naturally leading to a widespread word of mouth. The colours in the figure are proportional to the density of users. Noticeably, the most involved areas are North America, Australia, UK, and Scandinavia.

\begin{figure}[!ht]
    \centering
    \subfloat[]{
        \includegraphics[width=0.40\textwidth]{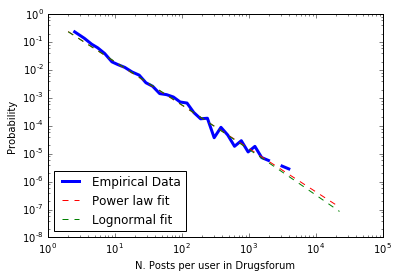}
        \label{fig:postperuserdf}
    }
    ~ 
    \subfloat[]{
        \includegraphics[width=0.40\textwidth]{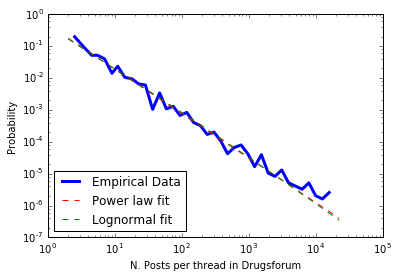}
        \label{fig:postperthreaddf}
    }
    \vfill
    \subfloat[]{
        \includegraphics[width=0.40\textwidth]{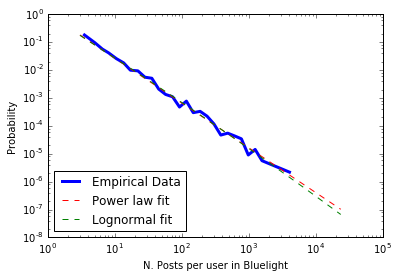}
        \label{fig:postperuserbl}
    }
    ~
    \subfloat[]{
        \includegraphics[width=0.40\textwidth]{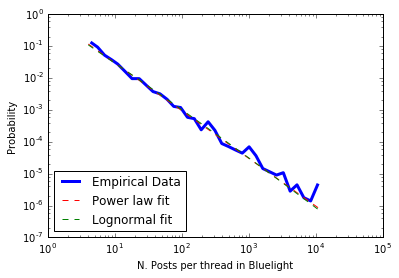}
        \label{fig:postperthreadbl}
    }
    \caption{Probability Density Function of the real data and the different distributions}\label{fig:fitting}
\end{figure}

\begin{figure*}[!htb]
    \minipage[t]{0.49\textwidth}
      \includegraphics[width=\linewidth]{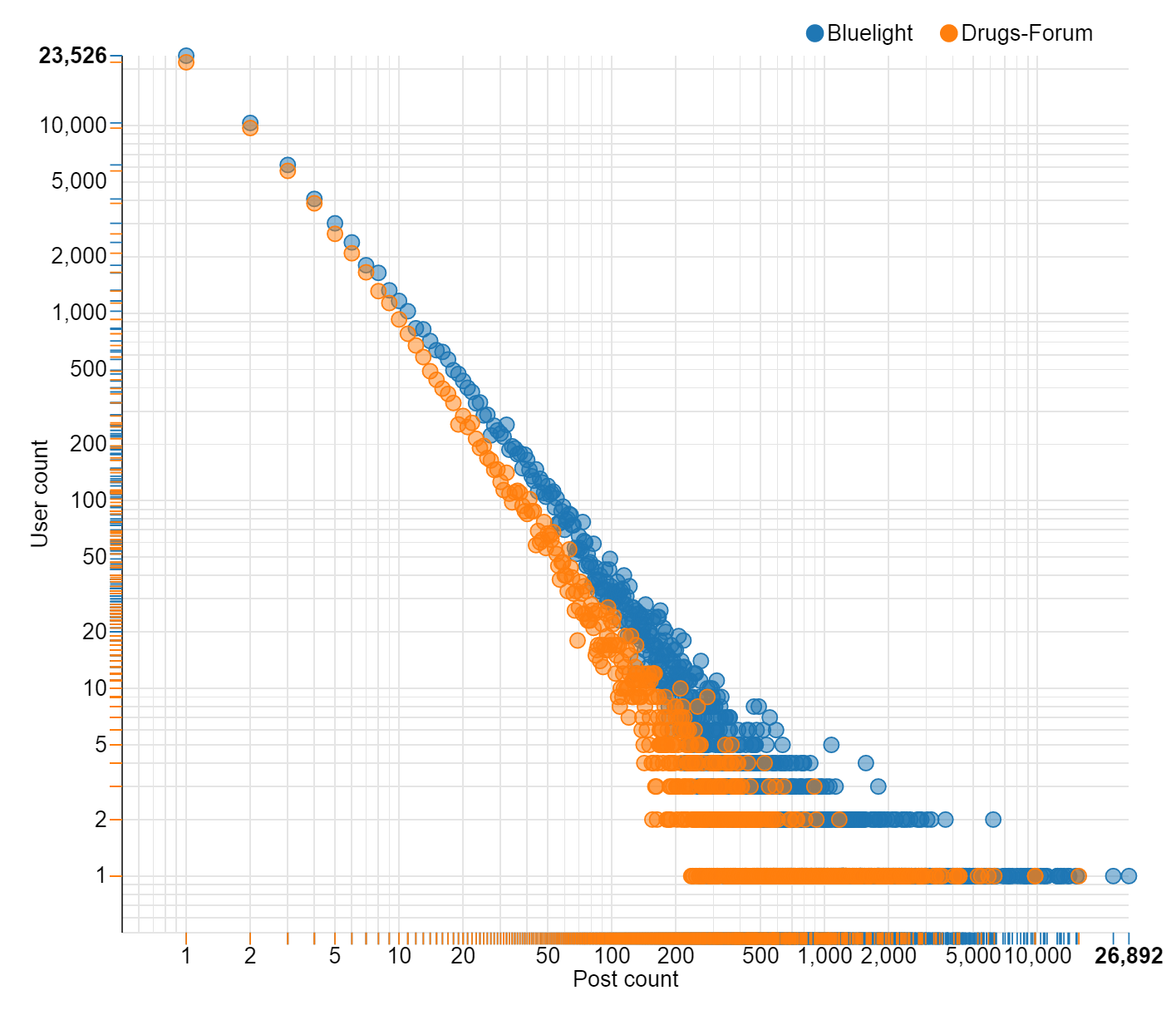}
      \caption{Posts per user.}\label{fig:postxuser}
    \endminipage\hfill
    \minipage[t]{0.49\textwidth}
      \includegraphics[width=\linewidth]{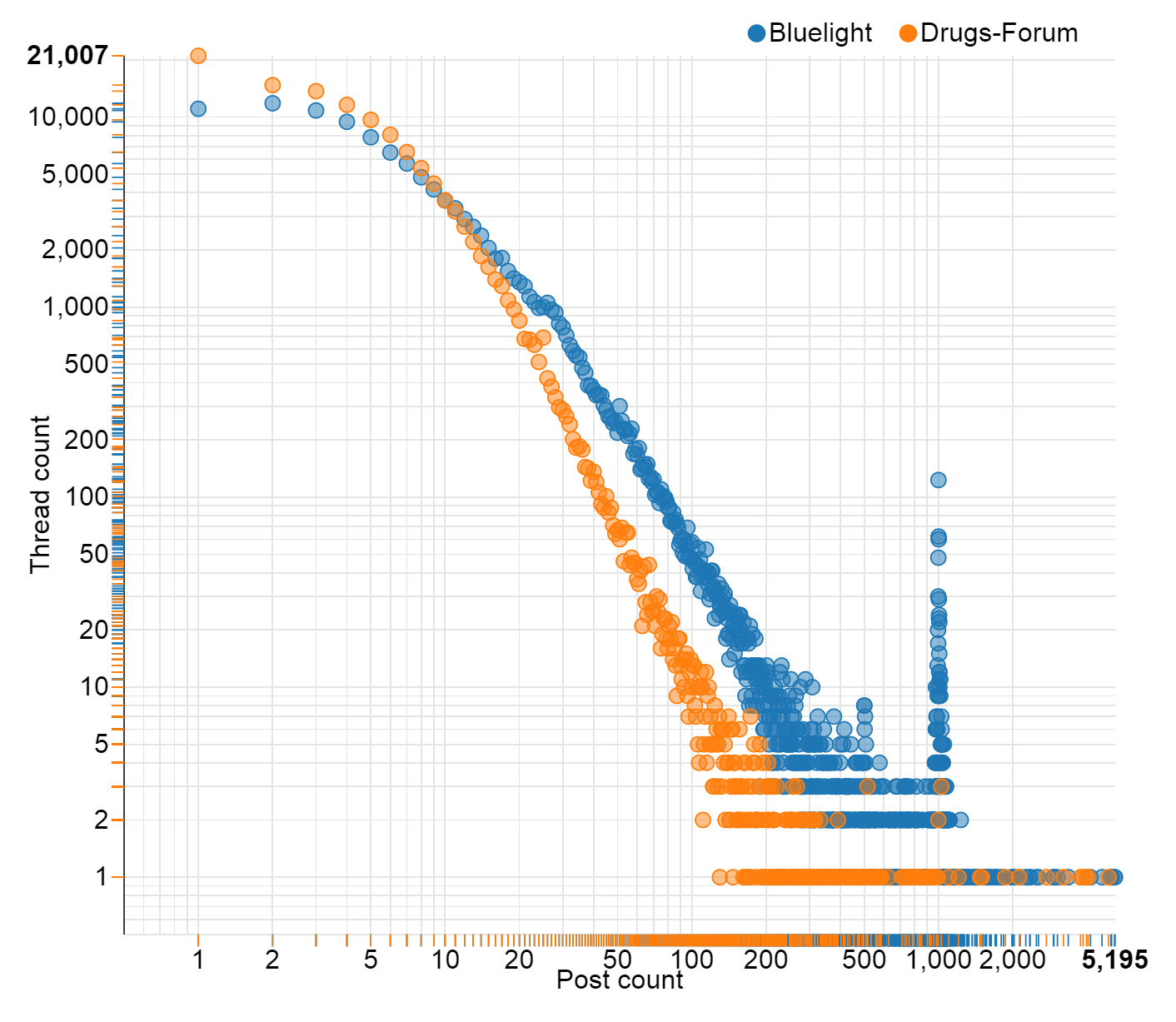}
      \caption{Posts per thread.}\label{fig:postxthread}
    \endminipage\hfill
\end{figure*}

We have also investigated some topological aspects of the forum, like
the number of posts per user and the number of posts per thread, on both forums (Figure~\ref{fig:postxuser}). With the \textit{powerlaw} Python package~\cite{Alstott2014}, we have compared the four real data distributions with the exponential, power law, truncated power law and lognormal distributions. The tool measured the {\it xmin}, no more than 4 for all the cases. Furthermore, with a {\it p}-value less than $10^{-8}$ for all the distributions, the power law distribution results in a better fit than the exponential one, as expected~\cite{Muchnik2013}. With regard to the lognormal and truncated power law distributions, the lognormal distribution fits slightly better than the power law one, while the truncated power law distribution fits better than the lognormal one. We can conclude that the (truncated) power law distribution assumption holds, as shown in~\cref{fig:postperuserdf,fig:postperthreaddf,fig:postperuserbl,fig:postperthreadbl}. These results highlight that there is a small amount of users responsible for most of the activity, on both forums.

It is worth noting that, even if Bluelight has about 0.6 times the number of users Drugsforum has (see Table~\ref{tab:forums}), the number of active users (i.e., that have written at least one post) is almost the same for both.
As for the distribution of posts per thread, shown in Figure~\ref{fig:postxthread}, Bluelight features a large number of threads having 1,000 posts. This is due to a limit on the maximum number of posts for certain threads: when exceeding the threshold, the moderators start a new thread for the discussion.

\subsection{Content analysis}
A text analysis that is really useful in our scenario is the measurement of volumes of discussion over time, given a term.
This investigation helps determining whether some drugs raise in popularity 
and in which section of the forum this happens, possibly obtaining some clues about the nature of the substance (being it a NPS or not).

\begin{figure}[!ht]
  \centering
  \includegraphics[width=0.8\textwidth]{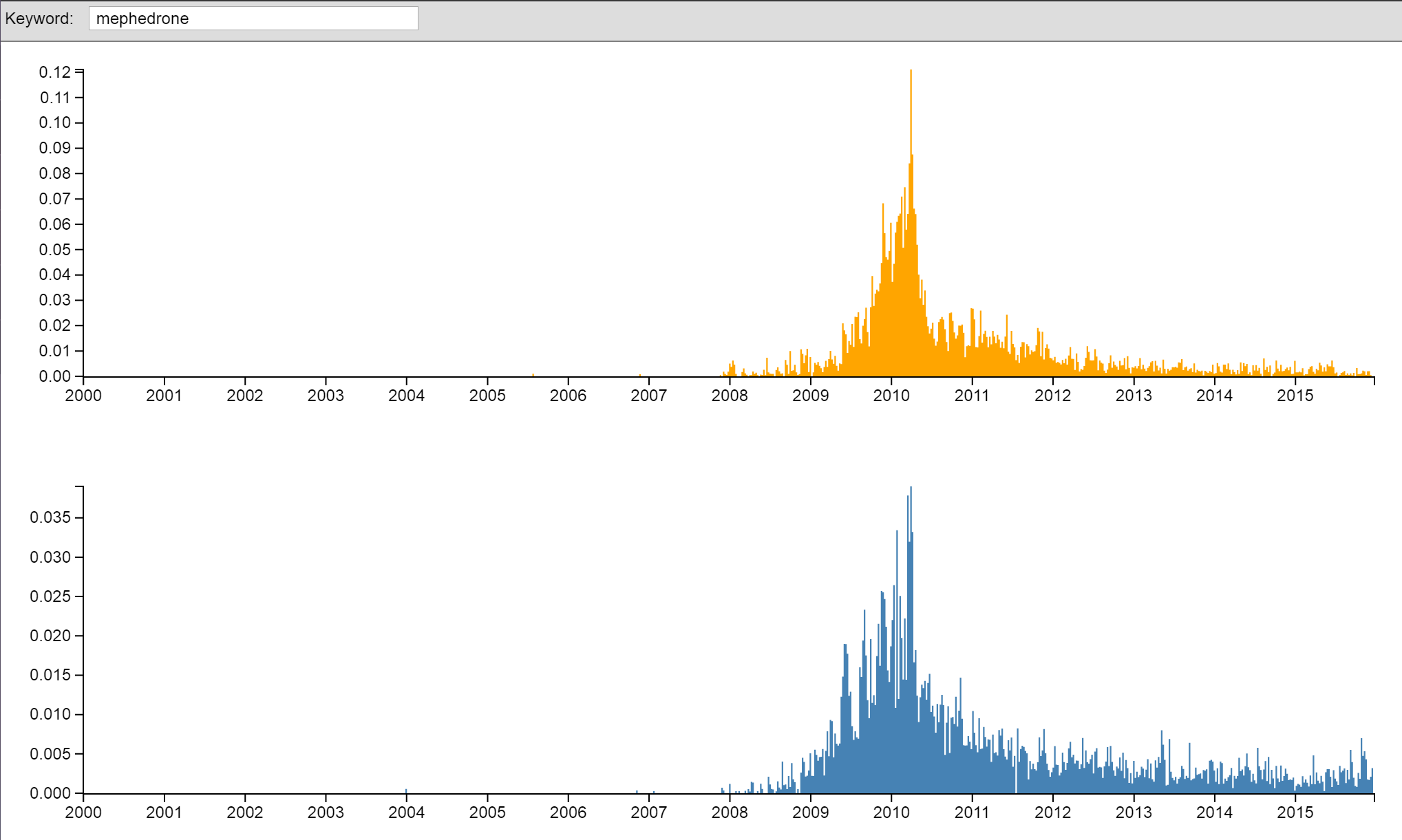}
  \caption{Frequency of ``mephedrone" over time, normalised to the whole volume of discussion, for Drugsforum (top) and Bluelight (bottom).}
  \label{fig:tf}
\end{figure}

Figure~\ref{fig:tf} shows the frequency of  the term ``mephedrone" over time, normalised to the whole volume of discussion, for Drugsforum (top) and Bluelight (bottom). Even if not identical, the shapes of the spike are similar, meaning that the substance has gained popularity within both the communities approximately at the same time.


\begin{figure}[!ht]
  \centering
  \includegraphics[width=0.8\textwidth]{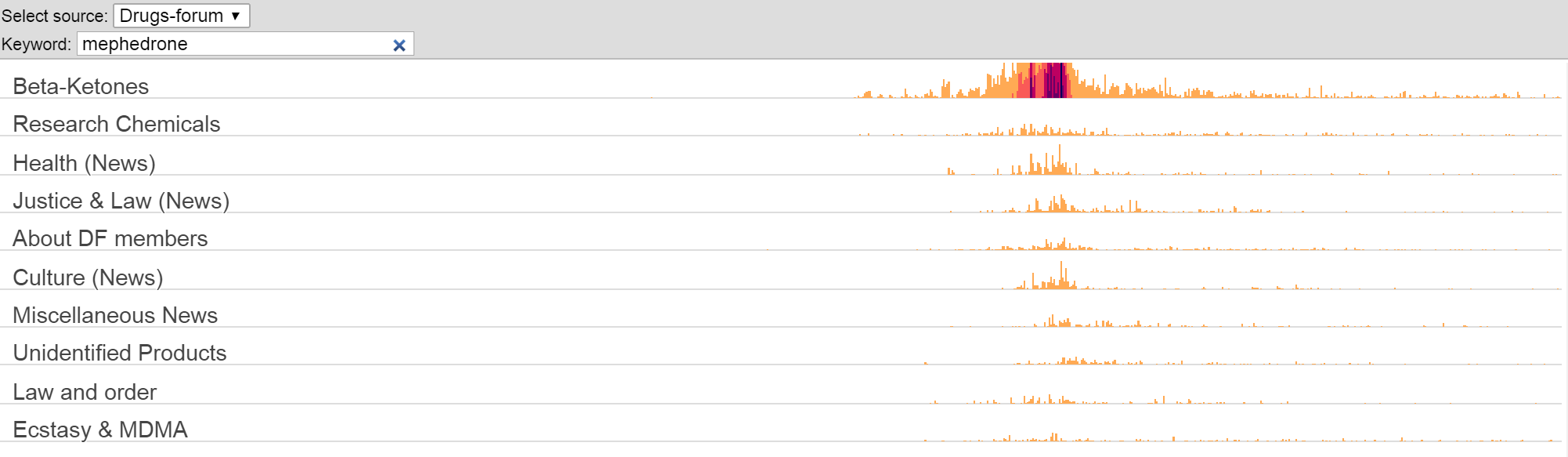}
  \caption{Horizon charts showing the frequency of a given term over time, for each subsection of the chosen forum.}
  \label{fig:horizon}
\end{figure}

Figure~\ref{fig:horizon} shows a higher level of detail: each line represents a subsection of the forum. As shown in the top-left part of the screenshot, we can choose which forum to analyse. A darker colour indicates a higher frequency of the term, for the corresponding time frame. The search for ``mephedrone" in Drugsforum shows a high volume of discussion in the first half of 2010 in a series of subsections, particularly in the one called ``Beta-Ketones". This indicates the category of the substance.

\begin{figure*}[!htb]
    \minipage[t]{0.49\textwidth}
      \includegraphics[width=\linewidth]{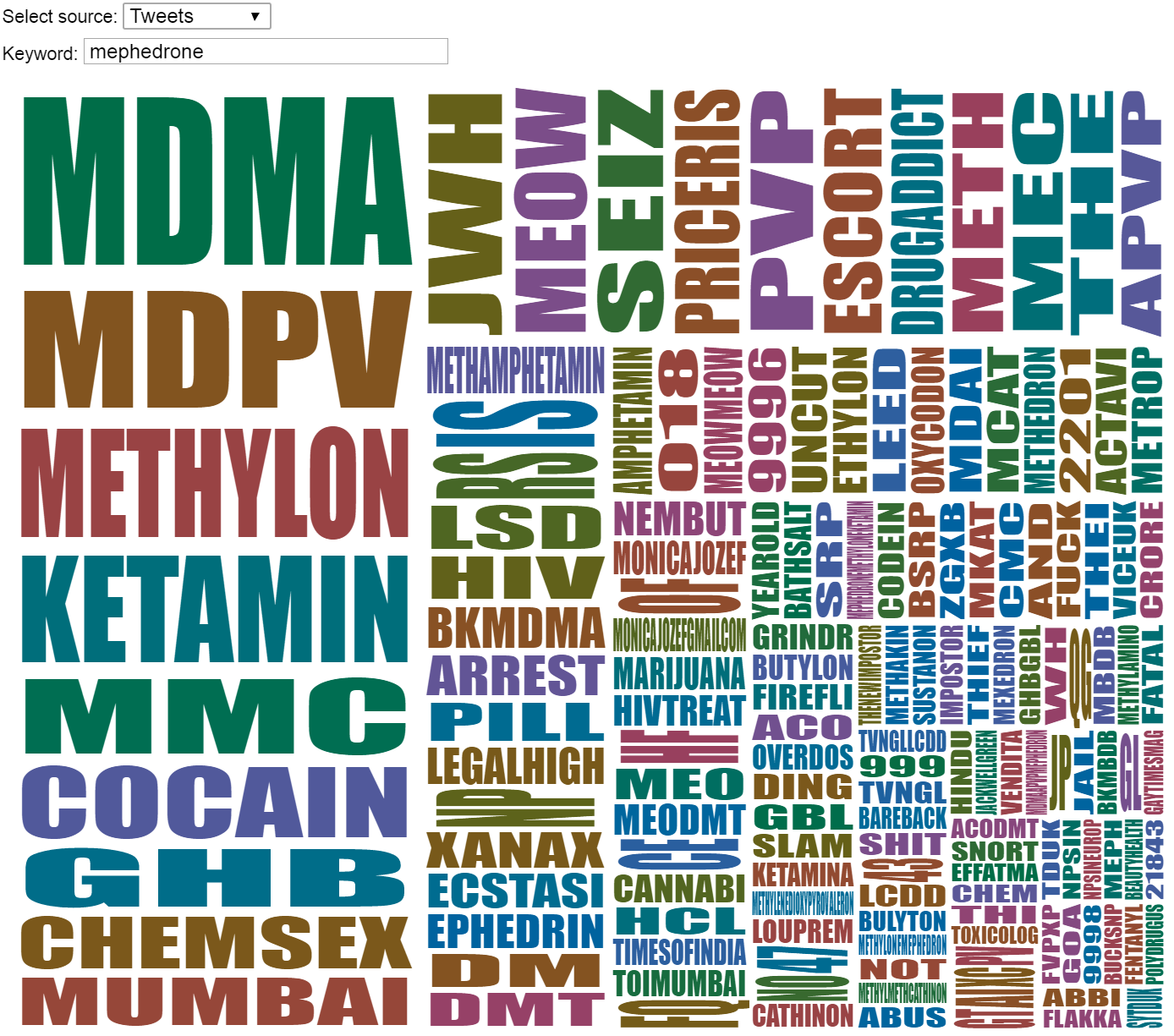}
      \caption{Zoomable wordcloud showing the most frequent terms co-occurring with ``mephedrone" in the Twitter dataset.}
  \label{fig:cloud}
    \endminipage\hfill
    \minipage[t]{0.49\textwidth}
      \includegraphics[width=\linewidth]{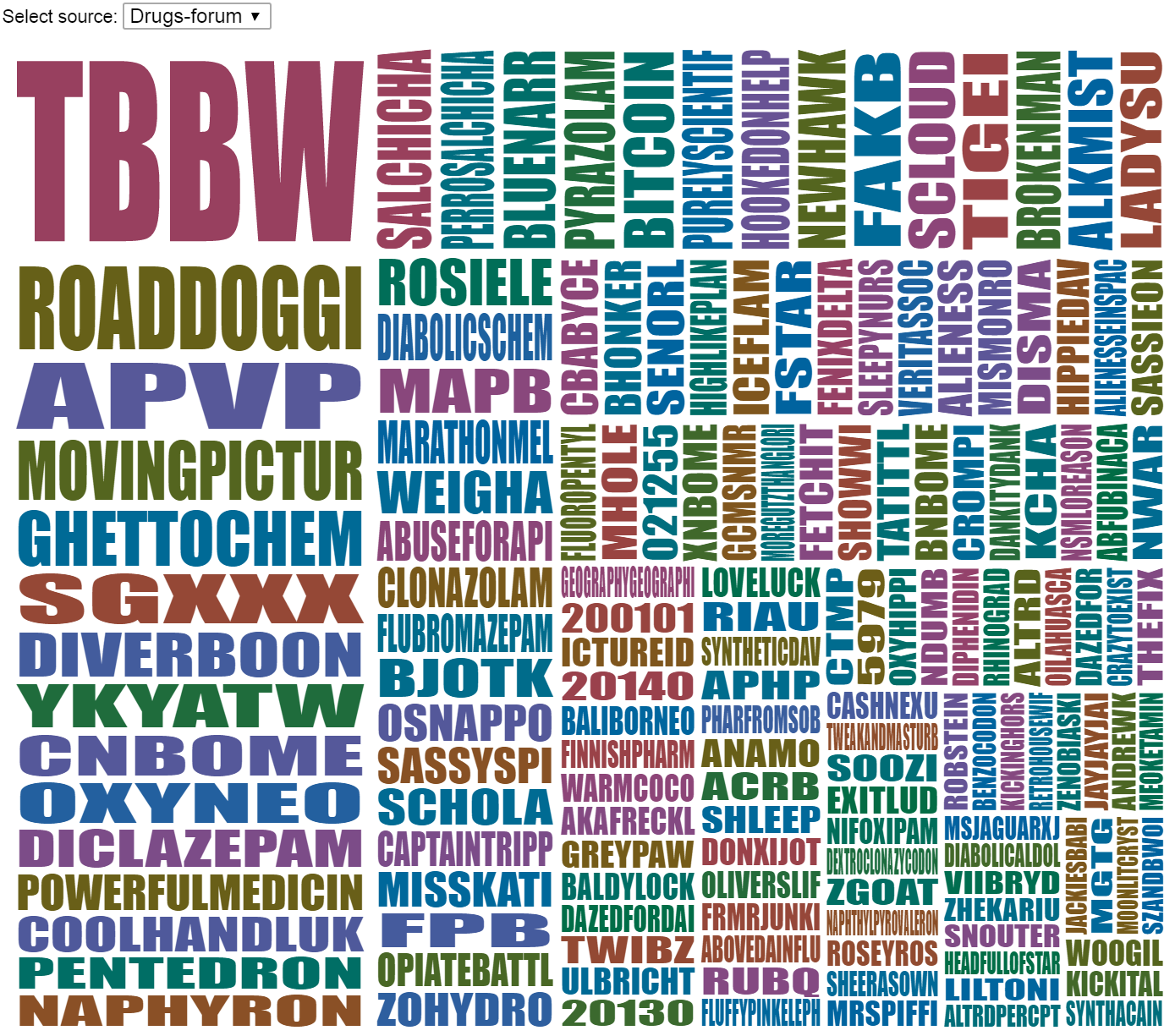}
      \caption{Zoomable wordcloud showing new terms in Drugsforum after 2010.}
  \label{fig:cloud2}
    \endminipage\hfill
\end{figure*}

As shown in the example of~Figure \ref{fig:cloud}, computing the terms that co-occur with a given one gives interesting insights. 
Indeed, the generated wordclouds may provide knowledge on substances that are similar, with similar effects and market trends.  In the figure, each word occupies an area that is proportional to its frequency. The wordclouds can be generated for both Twitter and the two forums.

Really endorsed drugs are presented and discussed in forums. To timely detect NPS, we have investigated neologisms and terminology on both the forums,  to discover new names. As an example, in Figure~\ref{fig:cloud2}, we plot the Drugsforum terms that appeared only after 2010. The result clearly indicates a lot of new drugs, appeared on the market from 2010 to 2015. It is possible to notice the name of some new drugs and medicines, such as $\alpha$-PVP, Diclazepam, Pentedrone, Naphyrone.

\subsection{NPS trading}
As a final set of analyses, we have explored the hyperlinks on the forums. Then, we have compared them with a comprehensive list of NPS online shops and with the links in the posts of monitored Twitter accounts. Not surprisingly, they do not overlap, meaning that forums discussions do not link shops. This is mainly due to the specific policies of the forums. 
We have also tested which are the NPS sold on the shops and also mentioned on forums, finding that almost every substance is mentioned. It is not possible to estimate the trade volume of NPS from online shops, but we can try to infer some information about popularity by observing the discussions in forums.
Checking the words frequency, we concluded that the very same substances are also advertised through Twitter. Table~\ref{tab:substances} reports an excerpt of some substances, with a measure of the discussion activity about them on Twitter, on forums and on online shops. In the table, the numbers in the column of online shops are the IDs of the shops, as in Table~\ref{tab:shops}. The meaning is: the drug is mentioned on those shops.  
\begin{table}[ht]
    \centering
    \begin{tabular}{|c|c|c|c|c|c|}
        \hline
        \bf{Drug} & \bf{Tweets} & \bf{Post BL} & \bf{Post DF} & \bf{Online shops} & \bf{First seen}\\
        \hline\hline
        MDAI & 913 & 3507 & 775 & 1, 3, 4, 9 & Bluelight\\
        \hline
        MDPV & 791 & 11304 & 3631 & 9 & Drugsforum\\
        \hline
        Methylone & 679	& 8254 & 5116 & 9 & Bluelight\\
        \hline
        AB-CHMINACA	& 584 & 16 & 33 & 4, 6, 9 & Drugsforum\\
        \hline
        Methiopropamine	& 515 & 329	& 232 & 2, 3, 7, 8, 9, 10 & Bluelight\\
        \hline
        1P-LSD & 483 & 612 & 69 & 1, 2, 3, 4, 9 & Bluelight\\
        \hline
        Etizolam & 1592 & 8629 & 2630 & 2, 4, 9 & Bluelight\\
        \hline
        Ethylphenidate & 965 & 2502 & 1268 & 2, 7, 9 & Bluelight\\
        \hline
        Synthacaine	& 217 & 124 & 60 & 3, 4, 9, 10 & Drugsforum\\
        \hline
        Diphenidine	& 193 & 779 & 80 & 2, 3, 4, 9 & Bluelight\\
        \hline
        Mexedrone & 39 & 113 & 14 & 1, 2, 3, 4, 9, 10 & Bluelight\\
        \hline
    \end{tabular}
    \caption{An excerpt of monitored substances, with no. tweets, posts and shops. Bluelight (BL) and Drugsforum (DF) are the two forums analyzed in this work. The last column highlights the forum where the substance has appeared first.}
    \label{tab:substances}
\end{table}

\section{Conclusions}
\label{sec:conc}
Today, New Psychoactive Substances (NPS) lie on a grey area, not precisely addressed by current regulations. NPS rapidly appear on -  and suddenly disappear from -  the market, with a consistent and continuous introduction of new surrogates, which leaves few margin for intervention by healthcare institutions and governments. 
This paper has put in evidence some unique features of online NPS forums and shops. Monitoring such websites and elaborating the available data made it possible to explore a large quantity of information, also across platforms, allowing analysts to perform comparisons among them.
We also gave a measurement of the relevance of NPS diffusion and advertisement, as well as user engagement. Furthermore, we showed how trading and discussions are correlated, through terms used by both online shops, social media, and forums, despite the prohibition, which hold on forums, to post explicit links to shops.
Noticeably, co-occurrences analysis and temporal analysis of neologisms are a valid support for NPS detection.

Currently, the analyses are led by the data scientist, which is assisted by the developed software. The analyses are applicable both to offline datasets and online streaming sources. We aim at fully automating some of the work,  
e.g., the detection of the psychic and physical effects of NPS on the individual, based on comments by the online users. 
Finally, we plan to extend the analysis to the dark web marketplaces.
\section{Acknowledgements}
\label{sec:agreements}
This publication arises from the project CASSANDRA (Computer Assisted Solutions for Studying the Availability aNd DistRibution of novel psychoActive substances), which has received funding from the European Union under the ISEC programme: Prevention of and fight against crime [JUST2013 / ISEC / DRUGS / AG / 6414].

\bibliographystyle{splncs03.bst}
\bibliography{bibliography}

\begin{thebibliography}{10}
\providecommand{\url}[1]{\texttt{#1}}
\providecommand{\urlprefix}{URL }

\bibitem{Alstott2014}
Alstott, J., Bullmore, E., Plenz, D.: powerlaw: a {Python} package for analysis
  of heavy-tailed distributions. PloS one  9(1),  e85777 (2014)

\bibitem{deBono2014}
de~Bono, B., Grenon, P., Helvensteijn, M., Kok, J., Kokash, N.: Apinatomy:
  Towards multiscale views of human anatomy. In: Advances in Intelligent Data
  Analysis XIII, pp. 72--83. Springer (2014)

\bibitem{Buntain2015}
Buntain, C., Golbeck, J.: This is your {Twitter} on drugs: Any questions? In:
  24th World Wide Web Conference - Companion Volume. pp. 777--782. ACM (2015)

\bibitem{Davey2012}
Davey, Z., Schifano, F., Corazza, O., Deluca, P.: e-{Psychonauts}: Conducting
  research in online drug forum communities. Mental Health  21(4),  386--394
  (2012)

\bibitem{Davies2010}
Davies, S., et~al.: {Purchasing ‘legal highs’ on the Internet —- is there
  consistency in what you get?} QJM  103(7),  489--493 (2010)

\bibitem{Freifeld2014}
Freifeld, C.C., Brownstein, J.S., Menone, C.M., Bao, W., Filice, R., Kass-Hout,
  T., Dasgupta, N.: Digital drug safety surveillance: monitoring pharmaceutical
  products in {Twitter}. Drug Safety  37(5),  343--350 (2014)

\bibitem{Hanson2013b}
Hanson, C.L., Burton, S.H., Giraud-Carrier, C., West, J.H., Barnes, M.D.,
  Hansen, B.: Tweaking and tweeting: exploring {Twitter} for non medical use of
  a psycho-stimulant drug ({A}dderall) among college students. Journal of
  Medical Internet Research  15(4),  e62 (2013)

\bibitem{Hanson2013}
Hanson, C.L., et~al.: An exploration of social circles and prescription drug
  abuse through {Twitter}. Medical Internet Research  15(9) (2013)

\bibitem{Hardy2015}
Hardy, R.A., Norgaard, J.R.: Reputation in the {Internet} black market: an
  empirical and theoretical analysis of the {Deep Web}. Journal of
  Institutional Economics  FirstView Article,  1--25 (2015)

\bibitem{Hielscher2014}
Hielscher, T., Spiliopoulou, M., V{\"o}lzke, H., K{\"u}hn, J.P.: Mining
  longitudinal epidemiological data to understand a reversible disorder. In:
  Advances in Intelligent Data Analysis XIII, pp. 120--130. Springer (2014)

\bibitem{Hillebrand2010}
Hillebrand, J., Olszewski, D., Sedefov, R.: Legal highs on the {Internet}.
  Substance Use \& Misuse  45(3),  330--340 (2010)

\bibitem{Inciardi2010}
Inciardi, J.A., Surratt, H.L., Cicero, T.J., Rosenblum, A., Ahwah, C., Bailey,
  J.E., Dart, R.C., Burke, J.J.: Prescription drugs purchased through the
  {Internet}: who are the end users? Drug and Alcohol Dependence  110(1),
  21--29 (2010)

\bibitem{Katsuki2015}
Katsuki, T., Mackey, T.K., Cuomo, R.: Establishing a link between prescription
  drug abuse and illicit online pharmacies: Analysis of {Twitter} data. Journal
  of Medical Internet Research  17(12) (2015)

\bibitem{Ledberg2015}
Ledberg, A.: The interest in eight new psychoactive substances before and after
  scheduling. Drug and Alcohol Dependence  152,  73 -- 78 (2015)

\bibitem{Muchnik2013}
Muchnik, L., Pei, S., Parra, L.C., Reis, S.D., Andrade~Jr, J.S., Havlin, S.,
  Makse, H.A.: Origins of power-law degree distribution in the heterogeneity of
  human activity in social networks. Scientific Reports  3 (2013)

\bibitem{Oconnor2014}
OConnor, K., Pimpalkhute, P., Nikfarjam, A., Ginn, R., Smith, K.L., Gonzalez,
  G.: Pharmacovigilance on {Twitter}? {M}ining tweets for adverse drug
  reactions. In: AMIA Annual Symposium. p. 924. American Medical Informatics
  Association (2014)

\bibitem{Paul2011}
Paul, M.J., Dredze, M.: You are what you tweet: Analyzing {Twitter} for public
  health. ICWSM  20,  265--272 (2011)

\bibitem{Pimpalkhute2014}
Pimpalkhute, P., Patki, A., Nikfarjam, A., Gonzalez, G.: Phonetic spelling
  filter for keyword selection in drug mention mining from social media. AMIA
  Summits on Translational Science p.~90 (2014)

\bibitem{Scott2015}
R.~Scott, K., Nelson, L., Meisel, Z., Perrone, J.: Opportunities for exploring
  and reducing prescription drug abuse through social media. Journal of
  Addictive Diseases  34(2-3),  178--184 (2015)

\bibitem{Schmidt2011}
Schmidt, M.M., Sharma, A., Schifano, F., Feinmann, C.: Legal highs on the
  net—{E}valuation of {UK}-based websites, products and product information.
  Forensic Science International  206(1),  92--97 (2011)

\bibitem{Soussan2014}
Soussan, C., Kjellgren, A.: Harm reduction and knowledge exchange---a
  qualitative analysis of drug-related {Internet} discussion forums. Harm
  Reduction Journal  11(1),  1--9 (2014)

\bibitem{Zawilska2015}
Zawilska, J.B., et~al.: Next generation of novel psychoactive substances on the
  horizon -- a complex problem to face. Drug and Alcohol Dependence  157,  1 --
  17 (2015)

\end{thebibliography}
\end{document}